# Generalized Insider Attack Detection Implementation using NetFlow Data

**Yash Samtani, Jesse Elwell**

Ridge High School, Perspecta Labs

Basking Ridge, NJ, U.S.A.

Emails: yashsamtani2@gmail.com, jelwell@perspectalabs.com

*Abstract*: Insider Attack Detection in commercial networks is a critical problem that does not have any good solutions at this current time. The problem is challenging due to the lack of visibility into live networks and a lack of a standard feature set to distinguish between different attacks. In this paper, we study an approach centered on using network data to identify attacks. Our work builds on unsupervised machine learning techniques such as One-Class SVM and bi-clustering as weak indicators of insider network attacks. We combine these techniques to limit the number of false positives to an acceptable level required for real-world deployments by using One-Class SVM to check for anomalies detected by the proposed Bi-clustering algorithm. We present a prototype implementation in Python and associated results for two different real-world representative data sets. We show that our approach is a promising tool for insider attack detection in realistic settings.

## I. Introduction

Insider attacks performed using networked hosts are an extremely important problem currently facing enterprise networks. Approaches to detect such attacks require solutions that are robust to different types of attacks that have never been seen, meaning that they must be good for general anomaly detection. Many previous signature-based and model-based techniques [6][7] rely on already constructed knowledge of different types of attacks and network protocols and are not general enough to use for anomaly detection of zero-day attacks. Instead of using static knowledge of known attacks, our proposed solution for anomaly detection combines weak indicators or features that demonstrate potential malicious or anomalous behavior using One-Class SVM and bi-clustering techniques [refs]. Our features use standard data available in Netflow schema such as inter-packet spacing and specific protocol information such as time for TCP SYNs to be acknowledged [ref]. Simple techniques such as using thresholds based on feature values or using only a few features results in high false positive rates. Instead, our approach of unsupervised machine learning combines many multi-dimensional features in a joint voting system by combining the outputs of the bi-clustering algorithm and One-Class SVM to further reduce some potential false positives created by weak individual indicators. We have demonstrated that our techniques achieve a detection rate of 94.9% and a false positive rate of 1.71%, when averaged between the two datasets we used, the UNSW-NB15 dataset [4], and the NSL-KDD dataset [5].

### A. Previous Work

Anomaly detection is a general process applied in many different areas, such as for cyber intrusion detection, credit card fraud detection, medical diagnosis and fault detection. Anomaly detection in communication networks is applied to one or more classes of observable data. These can be generally characterized into four classes: (1) sampled flow data (only taking some packets from flow data), (2) flow data (using 100% of flow data), (3) packet header data (breaks down more information about each packet), and (4) packet header and payload data (looks at carrying capacity of system packets are being transmitted over). As the granularity of observation increases for classes (1) to (4), so does the cost in terms of processing, storage and costs and generally do not lead to scalable solutions that can detect attacks in real time. Our observables fall into category (2). It is hard to define useful signatures over observables in classes (1) and (2), but, can serve as good data sources for weak indicators to be used for general anomaly detection.

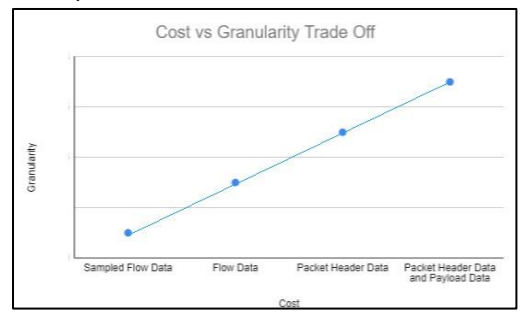

Fig. 1. Cost vs Granularity Trade Off in different types of flow data

Previous approaches [6][7] for anomaly detection based on network data have been plagued by high false positive rates of detection, which makes it impossible to deploy in a commercial or enterprise environment where false positives come at an extremely high cost to the enterprise deploying the detection system. Such techniques also have challenges with scalability due to sheer volume associated with network data.

Previous approaches [6][7] applied to this problem have been based on several different techniques - neural network algorithms [8], Bayesian networks [9], support vector machines (SVM) [10], rule-based techniques [11], K-nearest-neighbor [12] and several clustering-based techniques such as density-based spatial clustering (DBSCAN) [13]. Other techniques that have been applied to similar problems have been based on statistical parametric or non-parametric methods [14] and information-theoretic approaches [15]. These approaches varied in success, but all had common pitfalls of using relatively simplistic features that were not highly indicative of attack behavior. As a result, the performance of models trained using these features was not satisfactory and the models did not generalize well to attacks that are not included in the training data (as well as highlighted by the feature set used). This led to models that were relatively simple, and had high rates of classifying previously seen attacks, yet were plagued by a lack of generalizability as well as high false positive rates.

*B. Contributions*

We conducted ~~our~~ experiments using online NetFlow datasets, which ranged from real enterprise data collections to synthetic data representing a real world enterprise network environment. Our technique ~~exhibited~~ a high recall rate of true positives, as well as a low false positive rate. Also, while testing we found that many of the false positives detected by our system were to some degree anomalous data, meaning that it would be very hard to distinguish from any type of real attack. The two main contributions of our work were: (a) the implementation of the system and testing the utility of this system on publicly available dataset of real enterprise network data and (b) using features that are general to insider attacks, to enhance the ability of our model to detect zero day attacks.

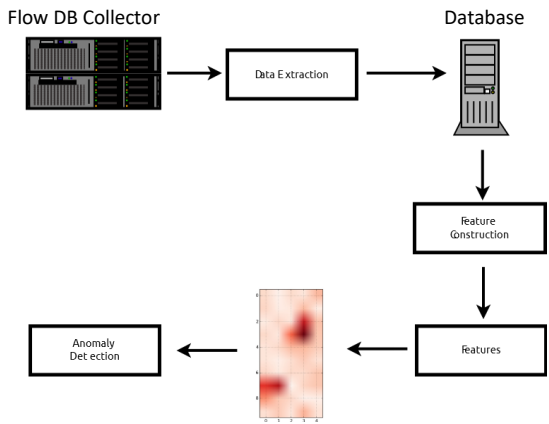

Fig 1. Anomaly Detection Architecture. ("Insider Attack Detection using Weak Indicators over Network Flow Data") We skip the Flow DB collection and data extraction phases in our setup, and instead work straight from publicly available datasets. We then use these datasets to come up with our feature sets in feature construction. Then, we use these features to pass through our model and detect possible anomalies in the system. We use parts of this diagram, the first 2-3 steps exist only in theory. (TOO LONG AS A CAPTION – MOVE TO THE TEXT)

II. ANOMALY DETECTION INFRASTRUCTURE

The infrastructure used for anomaly detection is shown above in Figure 1. We implemented different modules for extracting records from the NetFlow data, stored in a database and constructed features from the datasets to feed into our anomaly detectors.

A. *Datasets*

### 1) UNSW-NB15 Dataset

The UNSW-NB15 Dataset was the first dataset we used for network flow analysis. The dataset itself was generated in the Cyber Range Lab of the Australian Centre for Cyber Security (ACCS) and was generated to simulate both benign and attack data, similar to the data normally generated in the enterprise environment it was generated in. The attacks fell under nine categories, namely Fuzzers, Analysis, Backdoors, DoS, Exploits, Generic, Reconnaissance, Shellcode and Worms. The dataset had about a 3.4% attack rate (34 flows out of each 1,000 represented an attack), as well as about 2.5 million total pieces of data, making it an adequate for as a training model. Each piece of data is generated in the form of pcap files and has 49 features (including class/attack type labels), mainly flow and data traffic-based features, as well as information to identify and separate each flow. Flows were identified using a five tuple of Source IP, Destination IP, Source port, Destination port, and protocol type. When implemented in Python, we created a new type of object to store the five-tuple for each flow that was used to index and group records within the same flow. Some features such as total number of bytes were summarized using summation across multiple dimensions, for example, in each direction, duration, and total number of bytes over nonstandard/udp connections. For other features, a four-point summary was created for each flow, providing the mean, median, minimum, and maximum value of that feature throughout the flow. This allowed us to have more information about the flow, and also look across the timelines in a flow where an attack could've occurred. For example, there could be a flow that lasted 10 seconds in which there were 30 packets sent in one interaction that lasted a second, and 1 packet sent in the other 9. If you were measuring packets per second over the whole duration of the flow, you would get 3.44, but it doesn't give you the full picture of the spike in the flow. However, a statistic like maximum packets per second in one interaction, which would be 30 could give a much better picture of the packets sent and the spikes within data being transferred in the flow that could potentially indicate anomalous behaviors. This technique gave us many more unique features to feed into our anomaly detectors.

### 2) NSL-KDD Dataset

The NSL-KDD Dataset is an updated and improved version of the KDD'99 dataset, which is considered as the standard for intrusion detection. The original KDD dataset was derived from real US Air Force NetFlow data, with attacks peppered in, making it a very accurate representation of a real network environment, with anomalies that are not attacks being representative of real network noise. The data was collected using standard NetFlow collectors on each computer or unique IP address in the environment. The purpose of this data was to standardize a way to look at approaches for intrusion detection, which has made it a very commonly used dataset in the field. The NSL-KDD dataset updates got rid of highly similar data, which was redundant for machine learning algorithms, and reduced the amount of records in training and test sets to a more reasonable level. The dataset features four main types of attacks: Denial of Service, R2L (unauthorized access from a remote machine), U2R (unauthorized access to local superuser (root) privileges), probing (surveillance and other probing). These four types are then broken down into 38 distinct types of attacks, making it very hard to define useful signature for each attack, thereby requiring algorithms to find useful features for general anomaly detection. In addition, only 24 of the 38 types of attacks in the full dataset are shown in the training set (the dataset is already separated into training and test sets), making it test the critical ability of our system to be able to adapt to new attacks it hasn't already been exposed to. Moreover, the KDD dataset has a very high proportion of attack data of 46%, meaning that we need to lower it in order to test for a realistic environment. In achieve this, we got rid of attacks at random until we have the same 3.4% attack rate represented in the UNSW-NB15 Dataset, which makes comparing performance very easy. In terms of flow separation, the KDD dataset was much easier to work with than the UNSW dataset, as it is already separated into separate flows, with no need to manually index individual interactions. However, this also limited our ability to split features into different statistical metrics as we did with the UNSW-NB15 dataset. The dataset still has a fair amount of information about each flow, with 32 features being extracted from each flow, mostly traffic and connection-based features.

*C. Feature Set Construction*

A feature is a function that takes some category describing a flow, and turning it into one singular feature value to describe the value of that category on one flow. The values we either come up with mathematically using the feature values given in the datasets, or direct use of the feature values in the datasets become the feature values we use in our final feature set.

We cannot use every feature that is included in our dataset, as not every feature is informative to separating attack and benign data from each other, so we needed to differentiate informative features (weak indicators) from non-informative features that only either show in specific attacks or do not show in any attacks. In order to achieve this, we looked mainly at feature similarity between the features in the benign set or features in the attack set. We also looked at the features that were selected in random forests to classify attacks, as it uses self-error correction and eliminates features that are not useful to classify attacks from benign data. Lastly, we had to use some manual selection (trial and error), to see which features differentiated unique types of attacks, as some features had a lot of overlap in types of attacks detected.

Due to different types of features being collected, we had to use different features for the UNSW and KDD datasets. The features are listed here:

USNW-NB15: 1) Destination to Source Packets, 2) Minimum Source Inter-packet Spacing, 3) Minimum Destination Inter-packet Spacing, 4) Amount of bytes lost from Destination to Source, 5) Source to destination TTL value, 6) Time from the SYN_ACK to ACK packet in TCP connection setup, and 7) Source to Destination Packets per second

NSL-KDD Dataset: 1) Total Source to Destination bytes, 2) Total Destination to Source bytes, 3) Percentage of connections to different services, 4) Percentage of connections to same service, 5) Number of connections having the same destination port number, 6) Percentage of connections having the same Destination IP on the same service, 7) Percentage of erroneous connections having the same Destination IP, 8) Logged In status, 9) Percentage of erroneous connections having the same Destination Port, 10) Percentage of erroneous connections in the past two seconds to same destination host, 11) Number of connections in the past two seconds to same destination host.

For the Bi-clustering algorithm, either the (maximum value of the feature – value of a certain feature) or the reciprocal of the feature value is passed instead of the true feature value, as Bi-clustering relies on large feature values signifying anomalies in the data, we must normalize our feature values in the best way to make them useful to the bi-clustering algorithm. Once these features are extracted from the datasets, we also use normalization and scaling to put all our data on a similar scale. We use the standard normalizer and standard scaler given to us with the sci-kit learn package in Python to accomplish this, and we normalize slightly differently for the OCSVM and bi-clustering. For bi-clustering we use the standard l2 norm, and for bi-clustering we use the l1 norm, which keeps most feature values fairly low, while making anything higher than low values stand out more in the bi-clustering algorithm. This maintains a high variance of feature values of attack and benign data in our model, meaning that the feature values are able to indicate possible anomalies to our bi-clustering algorithm.

III. ANOMALY DETECTION USING BI-CLUSTERING

Bi-clustering is clustering the flows in a system between two different classes, here attack data and benign data, based on each class having similar feature values to other members of the same class. We create a graph to bi-cluster using our separated flows in each datasets and feature values.

As an example, Figure 2 is a graph in which each node on the left side of the graph is a flow of which data is collected, and each node on the right side is a feature, as described in Section II-C. In the graph, there is an edge connecting a node and a feature if the corresponding feature value is a non-zero value. In this specific example of a network, there are 10 flows and 5 features. The heatmap representing the state of the system is shown on the right of Figure 2, where each row in the heatmap is a flow corresponding to the number flow it is in the graph.

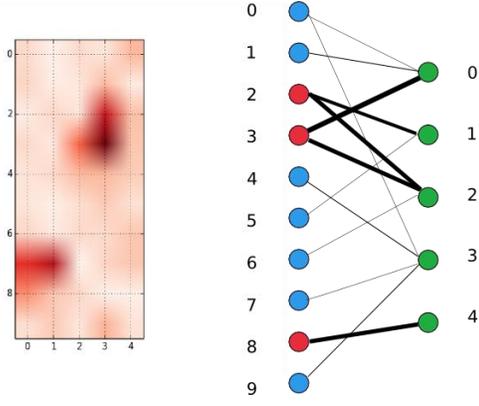

Fig. 2. Bi-clustering: from matrix to bi-graph representation ("Insider Attack Detection using Weak Indicators over Network Flow Data")

In the graph in Figure 2, the thickness of each edge is directly proportional to the value of the feature between the flow and its corresponding feature. In this graph, we can see that flows 2 and 3 are very closely connected to features 0, 1, and 2, making them a cluster in the overall scheme of the graph, as no other node has any similar strong connections to any of the same flows or features. Also, node 8 is strongly connected to feature 4, making nodes 8 and 4 another unique cluster in respect to the other nodes in the graph.

In general, the goal of the bi-clustering algorithm is to find groups of vertices that stand out from the rest of the graph, and to classify them as attacks based on their connectivity to other features.

### A. Bi-clustering Algorithm

For Bi-clustering, we will be given as an input a set of $n$ nodes or unique flows in the system described by $m$ attributes or features, the entire system can be represented as an $n \times m$ matrix $\mathbf{A} = \{a_{ij}\}$, where $a_{ij} \geq 0$ is the value of flow $i$ for feature $j$ in the system.

In a slightly more simple representation, this information can be shown as $G = (V_1 \cup V_2, E)$, G being a graph with edges E and vertices $V_1$ and $V_2$, with $V_1$ and $V_2$ being the set of unique flows in the network and the features, respectively, and E being the set of edges weighted by feature values. The endpoints of every edge connect one feature to one flow, meaning that the graph is a bipartite graph, as no two edges in either $V_1$ or $V_2$ are connected with each other, partitioning the graph into two sets of vertices.

The overall goal of bi-clustering is to take our matrix $\mathbf{A}$ and extract homogeneous blocks or sub-matrices from the matrix. We can say that the goal is to extract a set of flows with similar flows to label as anomalies from our inputted network flow. The best way of describing the problem that we have defined above is: given input of an undirected weighted bipartite graph $G = G(V,E)$, find the subgraph of $H^* \subseteq G$ that maximizes a certain criteria, or score function, $F(H^*)$. That is

$$H^* = \underset{H \in S}{\mathrm{argmax}} F(H) \quad (1)$$

where S is the set of all subsets of G.

The challenging part of creating the bi-clustering algorithm is to define our score function $F(H)$ so that our subset $H^* \subseteq G$ accurately captures a bi-cluster of graph G such that one cluster has a high frequency of anomalous data. Thus, we define the function $F(H)$ using a linkage function $\pi(i,H)$. The function $\pi(i,H)$ measures the connectivity of the flow $i$ to the rest of the nodes in subset H of the graph. Then, we $F(H)$ as the degree of association of the least connected node in H to the rest of nodes in H, i.e.,

$$F(H) = \min_{i \in H} \pi(i,H). \quad (2)$$

The definition of $F(H)$ makes it such that $H^*$ is a subset of nodes where the weakest degree of association between any two nodes in $H^*$ to the rest of the vertices in $H^*$ is the largest possible. For instance, if $\pi(i,H)$ is defined as the sum of the edge weights between vertex $i$ and all other adjacent vertexes in H, then the optimal H becomes one where each vertex in the optimal subset H is well connected to each other node in H.

In order to efficiently find a good solution for $H^*$, we need to make the score increase with each new vertex that is removed, i.e.,

$$\pi(i,H) \leq \pi(i, H \cup \{k\}), \forall i,k \in G. \quad (3)$$

Let us define the linkage function as:

$$\pi(i,H) = \sum_{j \in H, j \neq i} w_{ij} \quad (4)$$

where, $\pi(i,H)$ is the weighted degree of node $i$ in subgraph H.[1]

The linkage function in Eq. (4) is monotonically increasing, meaning that the minimum degree of the graph cannot increase by adding a new node to the graph, as the node would either have a lower degree and make the minimum decrease, or have higher and the minimum stays the same.

$$\pi(i, H \cup \{k\}) - \pi(i,H) \geq 0, \forall i,k,H. \quad (5)$$

Due to the monotonic decrease of $F(H)$ when removing nodes, we don't have to try every subgraph of G. Based on this, we can write an iterative algorithm to find the maximum value of the linkage function $F(H)$. We base our algorithm off the fact that the only way to obtain a subset with a value larger than $F(H)$, if such a value is possible to attain, is to delete the node with the minimum value of the linkage function, i.e.,

$$i^0 = \underset{i \in S}{\arg\min}\, \pi(i, H). \quad (6)$$

---
[1] In the following, edge degree is used with respect to this definition.

Due to the fact that $i^0$ has the minimum degree, any subset with a higher minimum degree clearly cannot include node $i$. Due to this, we now know that after we're given our graph G to begin with, we can eliminate the node of lowest degree again and again until we have our cut of the graph that maximizes F(H).

We derive an iterative algorithm for the above process in the pseudo-code as shown below.

*Pseudo-code of bi-clustering algorithm*

| |
|---|
| Input: bipartite graph G |
| Output: optimal subset H* |
| Initialization: F_best = 0, H* = empty_graph(), threshold, nodes = G.numNodes() |
| while G.numNodes()>(nodes*threshold) do |
|     node_min=min_degree(G) |
|     G.delete(node_min) |
|     F_score = min_degree(G) |
|     F_best = F_score |
|     H*=G |
| end while |

The subgraph H given by this algorithm is seen as the optimal bicluster for the graph G depending on the threshold value. The threshold represents what percentage of nodes in the graph should or should not be part of a bicluster and can be manually adjusted. This is mainly tuned by both training data, and can also be further tuned by inspecting trends when changing the threshold parameter.

*B. Anomaly Detection Using Bi-Clustering and One-Class SVM*

The bi-clustering algorithm from above is used as a classifier, and like any classifier, will give false positives and false negatives. The algorithm gives us no guarantees for the error to be within any certain percentage. In fact, a lot of the time the bi-clustering algorithm finds clusters in the graph, even when there are no anomalies. Therefore, although by the way the bi-cluster is set up, the highest scores usually correspond to anomalies, there may be slightly lower scores of benign data that form a cluster. Furthermore, the bi-clustering algorithm may create an output, even when there are truly no anomalies to classify.

In order to reduce and limit the false positive rate, we employed another classifier in order to filter out flows that don't show anomalous behavior. We combine these two classifiers with a form of a joint voting system, in which a flow will only be labeled as anomalous if both classifiers label it as an anomaly.

In network flow data, the ratio of benign data to attack or anomalous data is extremely high. This means that the task of classifying anomalies is in essence a one class problem, in which we need to make some bounds for one class, and classify any other data as anomalous, as opposed to training a classifier to recognize two distinct classes. Moreover, certain types of anomalies will not be seen in the training data. Due to these factors, many standard two-class machine learning algorithms won't work well to classify anomalous data from normal. Our idea was to use a one-class SVM, as it is ideal for one class problems with a small fraction of anomalies, as well as it having the ability to make bounds for normal data, without needing a lot of information about anomalies.

The idea of one-class SVM [13] is to minimize the space of the region that contain normal behavior data as shown in the following equation:

$$\min_{w, \xi_i, \rho} \frac{1}{2}||w||^2 - \rho + \frac{1}{\nu m}\sum_{i=1}^{n}\xi_i \quad (7)$$

The first two terms of Eq. (7) accounts for how big the volume of the region where normal samples lie is. The third term controls how large the penalty is for misclassifying. The parameter $\nu$ a prior set parameter and asymptotically represents:

- An upper limit on the percentage of anomalies that the SVM predicts.
- A lower limit on the percent of samples that will be used for constructing the decision boundaries.

These two facts only hold asymptotically, but they still provide some value on how to tune the parameter. In our settings, we set to 0.035, which is a little higher than the proportion of attacks in both datasets. Theoretically, this would limit the false positive rate to about 0.1%, but due to the facts only holding asymptotically we expect the actual false positive rate of the SVM alone to be a couple percent higher. The One-Class SVM outputs a list of unique flows as well as a label, labelling it as an anomaly or not.

The goal in using a One-Class SVM was to use it in conjunction with our bi-clustering algorithm, and labelling data perceived as both to be anomalous to be anomalous data. Therefore, only the hosts detected by both algorithms with a $\nu$ parameter of at least $0.035$ for SVM and a threshold value of $0.055$ for bi-clustering are be classified as anomalies[2].

IV. PERFORMANCE EVALUATION AND EXPERIMENTAL RESULTS

In order to determine the usefulness of our One-Class SVM and bi-clustering classifier, we need to test its performance on our datasets. We used test sets that were designed in such a way that it had a relatively low percentage of anomalies to keep the test set relatively realistic and designed them to look similar to our training data.

---

[2] These thresholds have been found by manual inspection. There is not restriction on the values of these two thresholds.

## A. Testing Design

We had to design a way to test our model which checked the ability of the classifier to accurately classify a wide degree of attacks, as well as classify new types of attacks as well. In the NSL-KDD dataset, data was already split into training and test sets, with training having 24 types of attacks, and the test set having 38 different types of attacks. We then take a random sample of the training and test sets to a reasonable size to which we can train the model with, and then test the model. This setup of taking random samples of the test set both allows us to get multiple points of data to avoid luck being a major factor in determining the effectivity of a model, as well as lets us test the robustness of the model to classify new unseen types of attacks, as we discussed being one of the major jobs of our model. For the UNSW dataset, data was not separated for us into training and test set. In order to train and test our model, we take a random sample of our data to use for each iteration, with a reasonable amount of data (we used 150,000 unique flows, this could be changed to either make accuracy better or make testing more efficient). We then split the first 75% of this random sample into training set, and the last 25% into test data, resulting in our training and testing set.

## B. Performance Evaluation of Combined One-Class SVM and Bi-Clustering

We tested each dataset five different times, the results of each test are displayed in the table below:

test and training test size are big enough to give us an accurate view. When we take averages of each dataset, we get that the USNW-NB15 dataset has an accuracy of 98.04%, recall of 98.65%, precision of 63.58%, false positive rate of 1.98%, and false negative rate of 1.27%. The NSL-KDD dataset has an overall accuracy of 98.29%, recall of 91.18%, precision of 70.25%, false positive rate of 1.44%, and false negative rate of 8.81%. As we can see, both have relatively similar results, but in the NSL-KDD dataset, both false positive rate and recall are considerably lower. Looking across the overall set of results, there is overall a very high recall in both systems, as well as a reasonable false positive rate lying between 1-2%, that leads us to believe that the combined system of the One-Class SVM and bi-clustering classifiers overall do a good job of classifying benign data from anomalous data. We could expect that in practice for the false positive rate to be considerably lower due to the much higher proportion of attacks to benign data than normal in the datasets. In both datasets, the percentage of flows that identified as an attack was about 3.5%, and we would expect that in a true enterprise environment, we would most probably have less than 1% of flows as true attacks, which would lead to a greatly diminished false positive rate.

## V. CONCLUSIONS AND FUTURE WORK

In this paper, we overall design and implementation of an anomaly detection technique using One-Class SVM and bi-clustering targeted at insider attack detection in an enterprise network environment. We tested our

| Dataset | Number of Flows | Number of Attacks | True Positives | False Negatives | False Positives | True Negatives | Accuracy | FP rate | Recall |
|---|---|---|---|---|---|---|---|---|---|
| USNW-NB15 | 37500 | 1259 | 1244 | 15 | 728 | 35513 | 98.02% | 2.00% | 98.81% |
| USNW-NB15 | 37500 | 1279 | 1257 | 12 | 716 | 35505 | 98.03% | 1.97% | 98.28% |
| USNW-NB15 | 37500 | 1283 | 1267 | 16 | 692 | 35525 | 98.11% | 1.91% | 98.75% |
| USNW-NB15 | 37500 | 1246 | 1234 | 12 | 726 | 35525 | 98.02% | 2.00% | 99.04% |
| USNW-NB15 | 37500 | 1288 | 1262 | 26 | 726 | 35486 | 98.00% | 2.00% | 97.98% |
| NSL-KDD | 17466 | 662 | 615 | 47 | 211 | 16593 | 98.52% | 1.21% | 92.90% |
| NSL-KDD | 17466 | 616 | 565 | 51 | 236 | 16614 | 98.35% | 1.35% | 91.72% |
| NSL-KDD | 17466 | 637 | 572 | 65 | 261 | 16568 | 98.13% | 1.49% | 89.80% |
| NSL-KDD | 17466 | 626 | 569 | 58 | 257 | 16582 | 98.20% | 1.47% | 90.75% |
| NSL-KDD | 17466 | 599 | 543 | 56 | 248 | 16619 | 98.26% | 1.42% | 90.65% |

TABLE I. PERFORMANCE EVALUATION OF ANOMALY DETECTION ALGORITHM (BI-CLUSTERING AND ONE-CLASS SVM COMBINED).

As we can see, performance between different tests does not to seem to have major differences, meaning that the implementation using the USNW-NB15 and NSL-KDD datasets, two of the most well-known intrusion detection datasets. We developed an anomaly detection system with input feeds as features representing weak indicators, and features that are good for detecting a wide range of attacks, as well as biclustering and one-class SVM based detectors, which demonstrated satisfactory results on our datasets. In addition, our approach is unsupervised, that is, it does not need any prior knowledge of the differences between normal and anomalous behavior. Our algorithm can be implemented as a system that is deployable as a real-world network-based insider attack detection system. As future work, we plan to add specific features relevant to attacks (blacklists, geographic location, refusing servers etc.) to further improve the performance of our model. Additionally, we plan to deploy this system on a live enterprise environment using inline feeds from enterprise routers and switches.